# Site selection for the 3.4m optical telescope of the Iranian National Observatory

Sadollah Nasiri[a*], Ahmad Darudi[b], Habib G. Khosroshahi[c] and Marc. Sarazin[d]

[a]Shahid Beheshti University, Faculty of Physics, Depatment. of Artronmy and Astrophysics, Tehran, Iran, 19839.

[b]University of Zanjan, Faculty of Science, Department of Physics, Zanjan, Iran, 45195-313.

[c]Institute for Research in Fundamental Sciences, School of Astronomy, Tehran, Iran, 19395-5531.

[d]European Southern Observatory, Karl-Schwarzschild-strasse 2, D-8046 Garching bei Munchen, Germany.

* Corresponding author

Sadollah Nasiri, nasiri@iasba.ac.ir

Ahmad Darudi, darudi@znu.ac.ir

Habib G. Khosroshahi, habib@ipm.ir

Marc. Sarazin, msarazin@eso.org

**Abstract.** Results of the site selection campaign conducted for the proposed 3.4m optical telescope of the Iranian National Observatory are reported. During the first 3 years, among 33 nominated regions throughout the country, the potential regions have been confined to four provinces, namely, Southern Khorasan located in the East, Kerman in the South East and Qom and Esfahan in the central part by examining the long-term meteorological and geographical parameters. Over the following 3 years, further astro-climate considerations and short-term atmospheric seeing measurements using differential image motion monitor (DIMM) technique, four candidate sites have been selected among these regions. Simultaneous seeing measurements were carried out on the four candidate sites and finally, the Dinava located at the common borders of the Esfahan and Qom provinces and Gargash in Kashan region were selected as promising sites. Continuous seeing measurements during 23 months yielded a median value of about 1.0" for both sites. The latitude and longitude of the Dinava are 50, 54 E and 34, 09 N, respectively and is situated at an altitude of about 3000 m and about 57 Km direct from Qom towards the South. The latitude and longitude of the Gargash are 51, 19 E and 33, 40 N, respectively and is situated at an altitude of about 3600 m and about 37km direct from Kashan towards the South. With two sites in hand, further comparative measurements of seeing and weather data continued at the two sites, which resulted in the selection of the Gargash site for the construction of the Iranian National Observatory.

## 1. Introduction

In 2000 the Astronomical Society of Iran initiated a plan for a national observatory, to host a medium size optical telescope which would promote research in observational astronomy. In this respect, preliminary studies concerning the long-term meteorology of candidate site parameters was begun. At the same time, a local workshop on instrumentation and observatory site selection was organized by inviting the experts from ESO, Nice University and Midi-Pyrenees Observatory in France, Indian Institute of Astronomy as well as from Iranian astronomical community. Meteorological studies on long-term data collected by the local Synoptic stations installed in or nearby regions of interest was undertaken as well as analyzing METEOSAT satellite data. The proposal for the Iranian National Observatory (INO) was prepared in 2003 and finally approved by the Iranian Ministry of Science, Research and Technology on March 2004.

Although the INO project is funded nationally, due attention has been given to international collaboration and a 3m-class-partnership in different stages of the project from development and design to construction and the planned science program.

Therefore, the telescopes and instrumentation are considered to be suitable facilities open to the international community.

Iran offers an attractive choice to host optical telescopes due to the dry climate of the central region, high mountains and the hour angle coverage, especially since there are no major alternative and equivalent facilities between the Indian subcontinent and Spain in northern hemisphere, which makes Iran especially suitable for the time-critical observations.

An astronomical site for an optical observatory is expected to fulfil several criteria including low atmospheric turbulence, as indicated by statistics of seeing and desired meteorological parameters such as wind speed and direction, cloudy and clear skies, air temperature and relative humidity, the perceptible water column, etc. Thus, the strategy adopted for this purpose was to perform a study based on existing meteorological and geographical data available from national databases to identify potential regions for inspection and short-term measurements of the seeing. Shortlisted sites were then studied for longer periods to obtain better characterization. The candidate regions, namely, Kashan, Marzi, Birjand and Kerman were selected among more than 30 potential regions around the entire country (see figure 1), (Nasiri & Abedini, 2003); Nasiri, 2003).

To determine the final best site we had to measure the local atmospheric turbulence, an important parameter, for these regions. We constructed four set of Differential Image Motion Monitor (DIMM), Rodier (1981), Sarazin & Rodier (1990), that were able to measure the statistics of the perturbations on the incoming wavefront, (Darudi & Nasiri, 2005). To produce a twin image the telescope was set slightly out of focus instead of using a wedge in front of one of the apertures. This technique is known as the defocused

DIMM (Tokovinin, 2002). All four DIMM were calibrated simultaneously and were taken to the four mentioned regions.

We developed software which made it possible to find the on-site seeing value by calculating the statistical variance of the relative motion of the centroids. By September 2004 our instrumentation was complete. Following an intense course on site selection for training the work force, we started to find the best site in each of the aforementioned regions. Finally, Kolahbarfi in Kashan, Dinava in Marzi, Mazarkahi in Birjand and Sardar in Kerman were selected as the principal candidate sites of each region. The simultaneous night time seeing data started from April 2005 in a continuous manner throughout the whole year. After 11 months the Birjand site was rejected compared to the other 3 sites. 4 months later the Kerman site was also rejected (Nasiri & Darudi, 2006). By March 2007 we stopped the seeing data collection and observed no crucial difference between Kolahbarfi and Dinova sites. Further seeing measurements were made on Gargash (a steep peak) located at the same region, but a few hundred meters higher in altitude, which yielded the best conditions. It was therefore selected as the final INO site.

In section 2 we describe the countrywide survey on the initial 3 dozen potential regions, mainly using meteorological and seismic data which led to a short list of the 4 candidate regions. The procedure followed to determine the best candidate site in each region is also discussed. In section 3, the seeing monitoring instruments, techniques and measurements are presented. Results and discussion are presented in section 4 and section 5 is devoted to conclusions.

**2. Countrywide survey**

The landscape in Iran is dominated by rugged mountain ranges, separating several plateaus from one another. The populous western and northern parts are the most mountainous, the Zagros and Alborz Mountains extending south-west and north-east, respectively. Damavand at 5,671 m is the highest peak in the country. The eastern and central regions are mostly uninhabited deserts.

In order to identify the potential regions, Meteosat cloud coverage data between 1983 and 1993 were studied and 33 regions across the country were identified. These regions are marked by black dots in Fig.1.

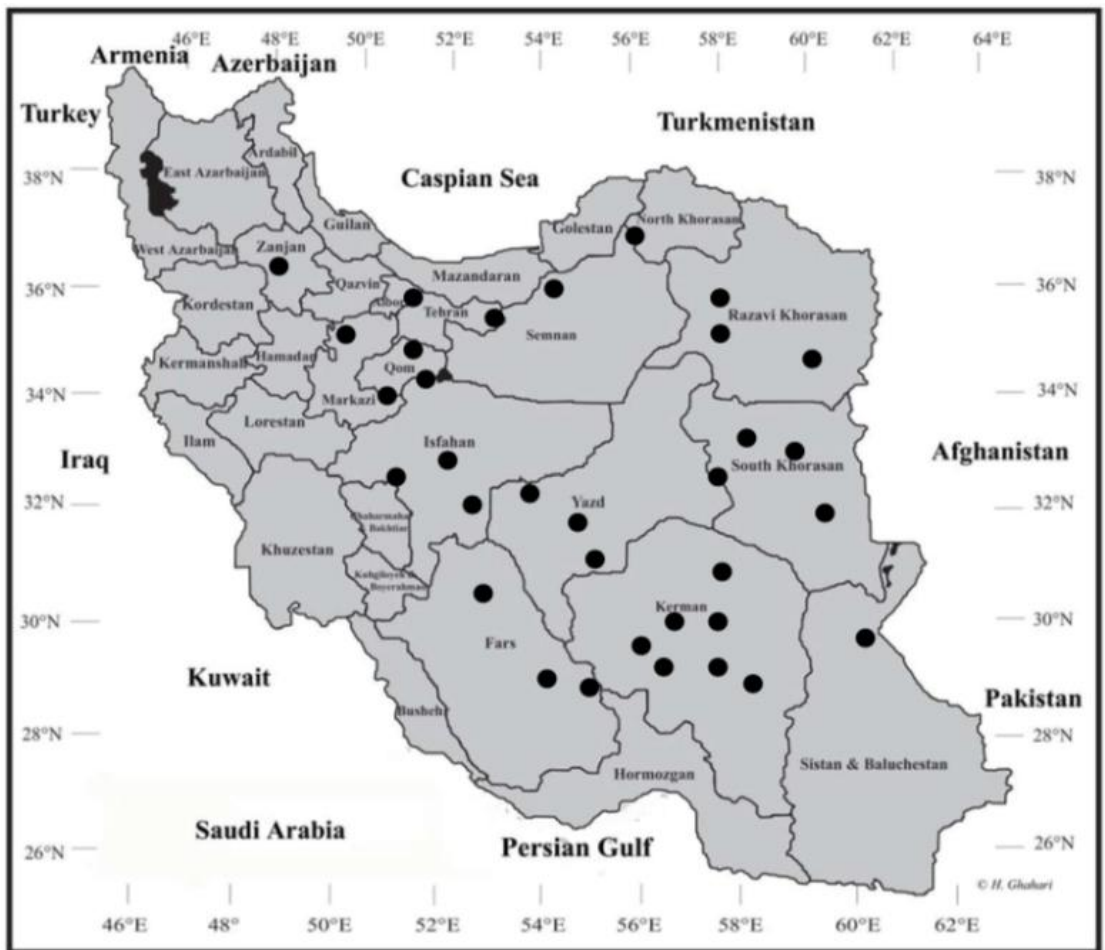

Fig.1) The potential regions in the country are marked by black circles

The study of sunny days indicated that the central desert area of Iran is less likely to be cloudy with more than 250 days per year of bright sunshine. Although this does not directly give a measure of clear nights, there is no indication that this trend can change significantly during the night. As far as the bulk cloud is concerned, the clouds in central region are those migrated from Black Sea and there are no major local sources which can change the trend between day and night. The northern coast of the Persian Gulf also benefits from a clear sky, however, there can be dust coming through the Arabian subcontinent, and also aerosols relating to the oil industry, the lack of high altitude mountains reduces their potential quality.

To create a short-list of the promising sites a study of the annual, seasonal and monthly wind-rose, as well as the percentage of the calm wind days was carried out over a period of 20-50 years using the data collected by the local Synoptic stations available in (or nearby) the regions of interest, (Nasiri & Abedini, 2003).

There are many active geological faults that give rise a relatively unstable seismic situation in Iran. We performed a study based on an eligible data recorded for the past available seismic events and the ground acceleration of the active faults, (Nabikhani, 2004; Nasiri 2004).

Combining the results of wind-rose studies, cloud coverage, seismic and faults activities and aerosol index for 33 nominated regions over the country, we reduced the potential regions to four provinces, namely, Southern Khorasan located in the East, Kerman in the South East and Qom and Esfahan in the central part of the country, Nasiri & Darudi (2005).

To find the best site at each region, four teams were equipped with a DIMM (Diffrential Image Motion Monitor) instrument and necessary portable equipment such as laptop computers, electricity generators, living quarters etc., were simultaneously mounted and operated on these regions. After a couple of months, the whole area in each of the four regions were visited and DIMM data were collected from local peaks during the night time. During this phase, about 6 sites in Qom Province, 8 in Kashan (located in Isfahan province), 21 in Kerman, and 7 in Birjand (located at the Southern Khorasan province) are examined, respectively. In addition to the seeing results, the other local parameters (see the next section) are compared and finally, 4 sites as shown in Table 1 were selected for further competence by long-term seeing measurements.

| Site name | Province | Latitude/Longitude | Altitude | Distance |
|---|---|---|---|---|
| Kolahbarfi | Esfahan | 51, 26', 44"/33, 37', 55" | 3200 m | 39 km direct (70 km by road) |
| Mazarkahi | Khorasan | 59, 29', 22"/32, 38', 28" | 2700 m | 32 km direct (55 km by road) |
| Sardar | Kerman | 57, 01", 24"/30, 34', 42" | 1400 m | 55 km direct (98km by road) |
| Marzi | Qom | 50, 55', 99"/34, 08', 24" | 3000 m | 53km direct (75 km by road) |

Table 1) 4 sites to be studied by further long-term seeing measurements

To enable continuous long-term seeing measurements on these candidate sites, some living structural facilities and road were constructed (see Figs. 2).

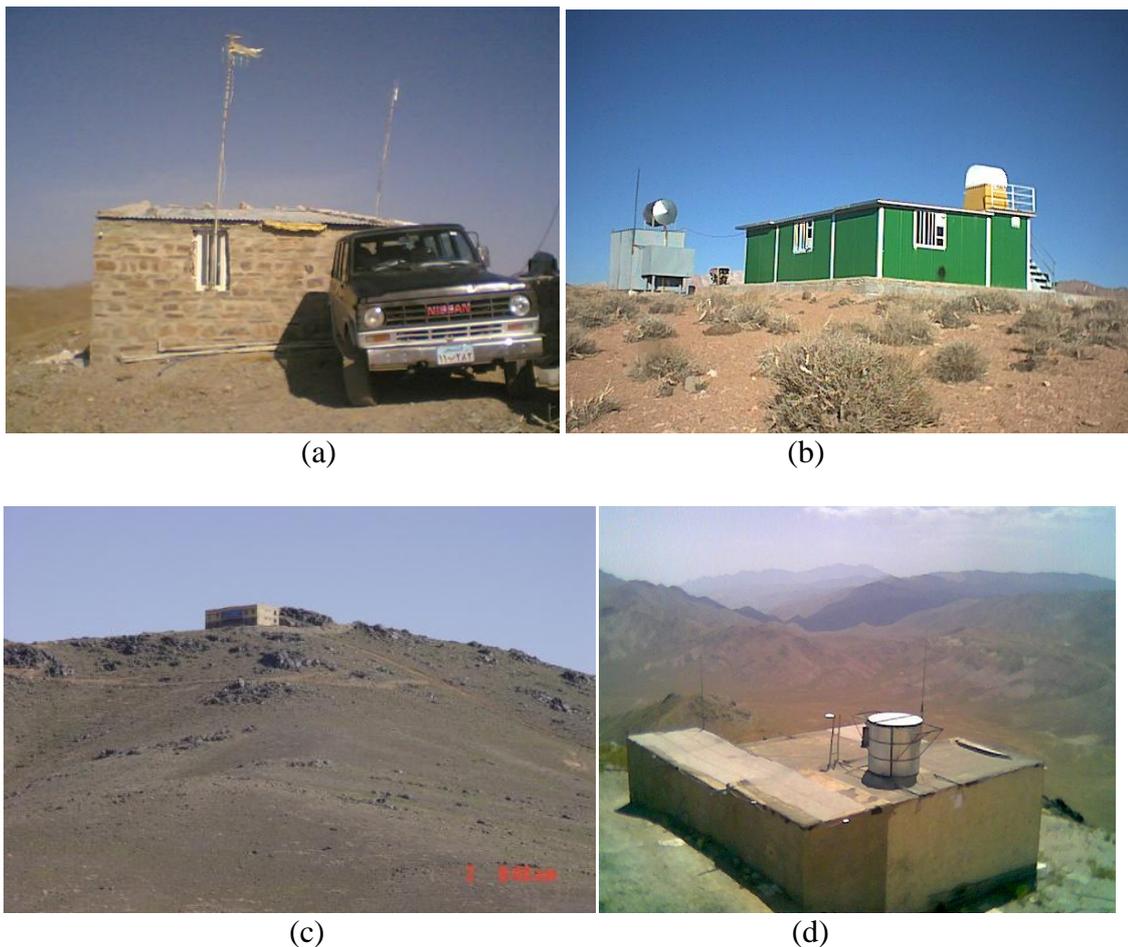

(a)      (b)

(c)      (d)

Fig.2) Views of a) Marzi site in Qom, b) Sardar site in Kerman, c) Mazarkahi site in Birjand and d) Kolahbarfi site in Kashan

## 3. Seeing measurements

Quantitative measurement of the image quality known as the seeing parameter was performed by a standard seeing monitor (DIMM). In DIMM method, one measures the wavefront slope differences over two small pupils some distance apart, independent of tracking errors and wind shake. In our DIMM (see Table 2 for its specifications) the starlight goes through two 8 cm circular sub-apertures cut in a mask and placed at the entrance of a F/10 Celestron 11" telescope equipped with GPS.

| CCD camera | ST2000 |
|---|---|
| CCD Pixel size | $7.4 \times 7.4 \mu m$ |
| No. of pixels | $1600 \times 1200$ |
| Pixel array | $11.8 \times 8.9\ mm$ |
| Full frame acquisition | 4.5 seconds |
| Telescope | Celestron 11 with GPS |
| Sub-aperture size (D) | $80\ mm$ |
| Sub- aperture separation (S) | $200 mm$ |

Table 2) INO DIMM specifications

The dual star image was obtained by setting the telescope slightly out of focus. The simulation which was completed using the Tokovinin method, (Tokovinin, 2002) yield but thin defocused technique introduced bias on the DIMM measurements. A simple method to control this bias is computed using the Strehl ratio for each spot. We rejected data having a Strehl ratio below 0.3, (Tokovinin, 2002). The CCD camera was set to grab images with 10ms exposure time while working on quarter image size mode. The data of 70 images recorded every 3 minutes was transferred to a laptop via USB port. The alignment and tracking the selected star were done automatically by the telescope equipped with GPS.

A software package then enacted the following steps on each image:

*i.* Before taking each set, the dark frame was grabbed and subtracted from image.

*ii.* A circular window was applied to each spot with a diameter slightly greater than the Strehl spot size made by each DIMM aperture.

*iii.* Any frame having a spot with a saturated pixel was rejected.

*iv.* For the reduction of the noise, the intensity of pixels below a threshold value was assumed to be zero. The number of nearby remaining pixels had to number at least 3, otherwise the frame was rejected.

*v.* The sky background was calculated and subtracted from image intensity.

*vi.* The center of intensity of each spot was calculated.

*vii.* For each spot the horizontal and vertical axes were computed. If the ratio of the minor and major axis became less than 0.4, the frame was rejected.

*viii.* The distance between two spots was calculated in longitudinal and transverse directions. For simplicity, before taking data, the CCD was rotated until the longitudinal direction of spots became along the horizontal direction of CCD pixel array.

*ix.* The standard deviation of the distance between the spots was calculated for each set in longitudinal ($\sigma_l$) and transverse ($\sigma_t$) directions.

*x.* To take into account the instrumental noise, the standard deviation of excess noise made by the instrument was measured in a laboratory. For this purpose a fixed point source was prepared and the standard variation of the image motion was calculated.

Using the Kolmogorov (1941) model the longitudinal and transverse values of the seeing parameter, denoted by $\varepsilon_l$ and $\varepsilon_t$, respectively, could be estimated by

$$\varepsilon_l = 0.976 \left(\frac{\lambda}{r_{0l}}\right) = 0.976 (\frac{D}{\lambda})^{0.2} [\frac{(\sigma_l^2 - \sigma_n^2)\cos\gamma}{k_l}]^{0.6}$$

$$\varepsilon_t = 0.976 \left(\frac{\lambda}{r_{0t}}\right) = 0.976 (\frac{D}{\lambda})^{0.2} [\frac{(\sigma_t^2 - \sigma_n^2)\cos\gamma}{k_t}]^{0.6}$$

Where

$$k_l = 0.364\left[1 - 0.523b^{\frac{-1}{3}} - 0.024b^{\frac{-7}{3}}\right],$$

$$k_t = 0.364\left[1 - 0.798b^{\frac{-1}{3}} - 0.018b^{\frac{-7}{3}}\right],$$

$$b = \frac{S}{D}.$$

$\gamma$ is zenith angle of observing star, $\lambda$ is the wavelength, $r_0$ is the Fried's parameter, $D$ is the Sub-aperture size and $S$ is the Sub-aperture separation. Finally the site quality defined by the long exposure Full Width at Half Maximum (FWHM expressed in arcsec) at the wavelength $\lambda = 0.5\mu m$ was obtained as the average of both $\varepsilon_l$ and $\varepsilon_t$

$$FWHM = \frac{1}{2}(\varepsilon_l + \varepsilon_t)$$

To have a simultaneous site testing measurements which allows a fair comparison of different candidate sites over a long time, four sets of DIMM mounted on four candidate sites collected the on-site seeing data continuously and simultaneously for about two years. Fig.3 shows the mean monthly seeing parameter for 23 consecutive months, May 2005 to April 2007, for four candidate sites.

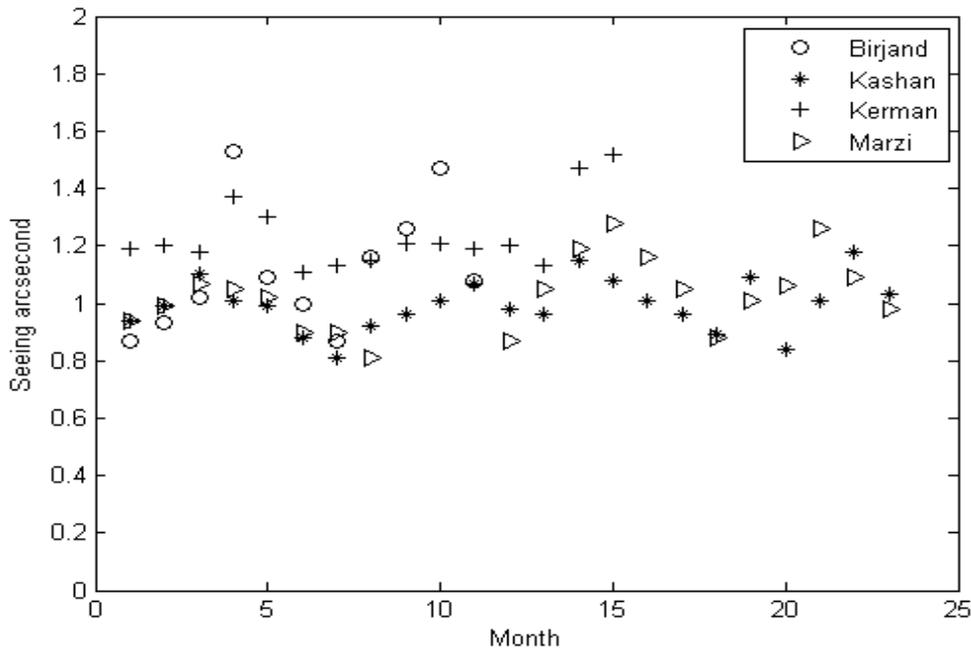

Fig.3) The seeing parameter of candidate sites from May 2005 to April 2007

As shown by the data in Fig.3, the Birjand site was rejected after 11 months, in March 2006. During this period the mean seeing of Birjand site was $1.2"\pm0.1$. The Kerman site was rejected after 15 months, in August 2006 as its mean seeing was $1.3"\pm0.1$ at this period. The seeing data of Kashan and Marzi sites were obtained from April 2005 to March 2007 and the seeing conditions of these sites were nearly similar and with median seeing of $1.0"\pm0.1$ as shown in Figs. 3 and 4. As seen in Fig. 3, it seems that there is some correlation between the seeing conditions of Marzi and Kashan, as there are two relatively close sites with nearly the same elevation.

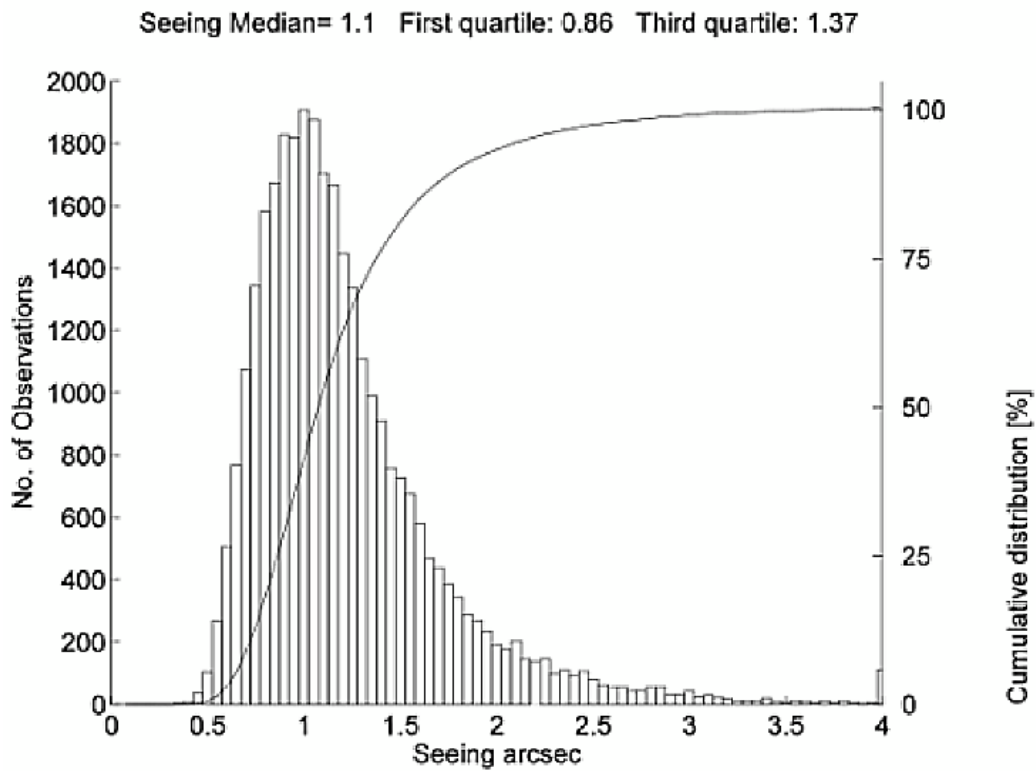

Fig.4a) Seeing histogram of Birjand site for 11 months.

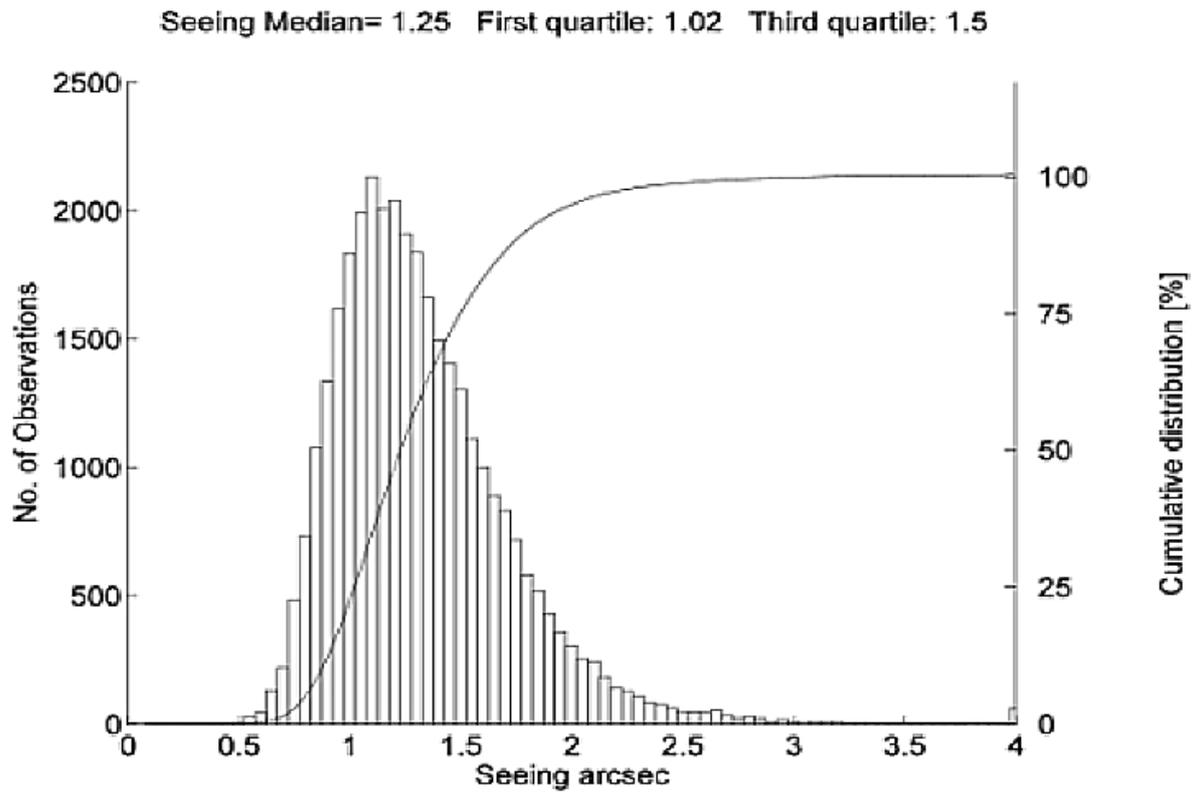

Fig.4b) Seeing histogram of Kerman site for 15 months.

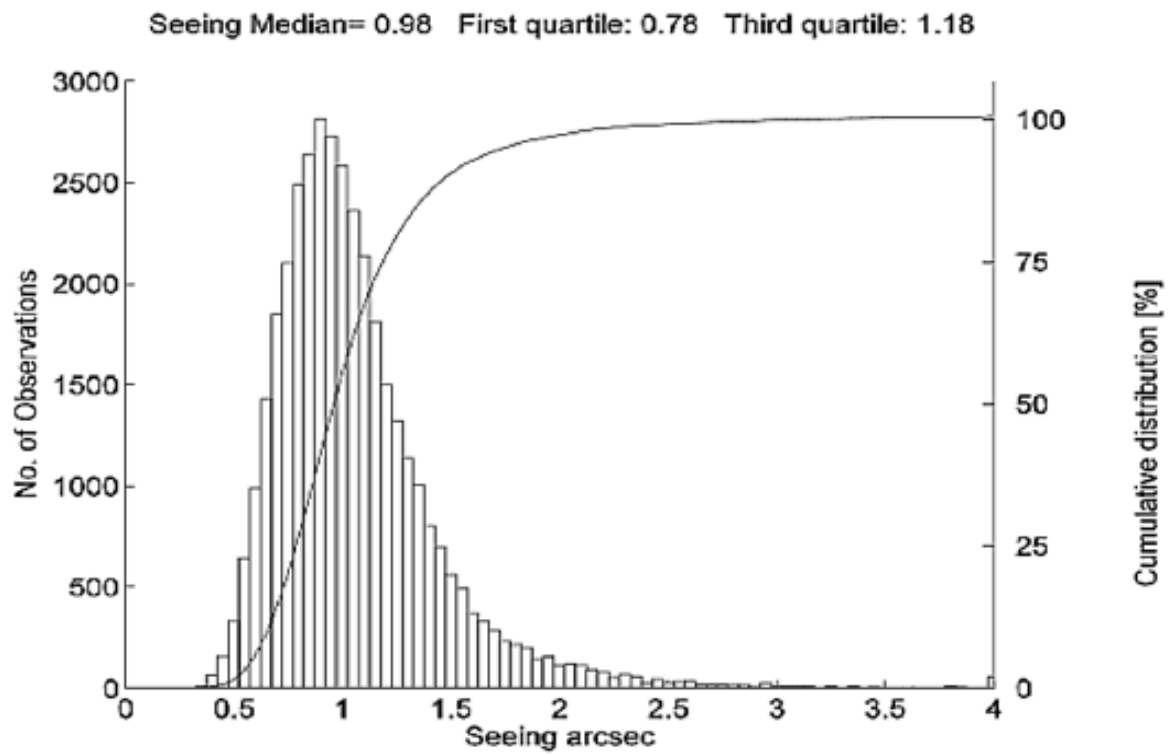

Fig.4c) Seeing histogram of Kashan site for 23 months.

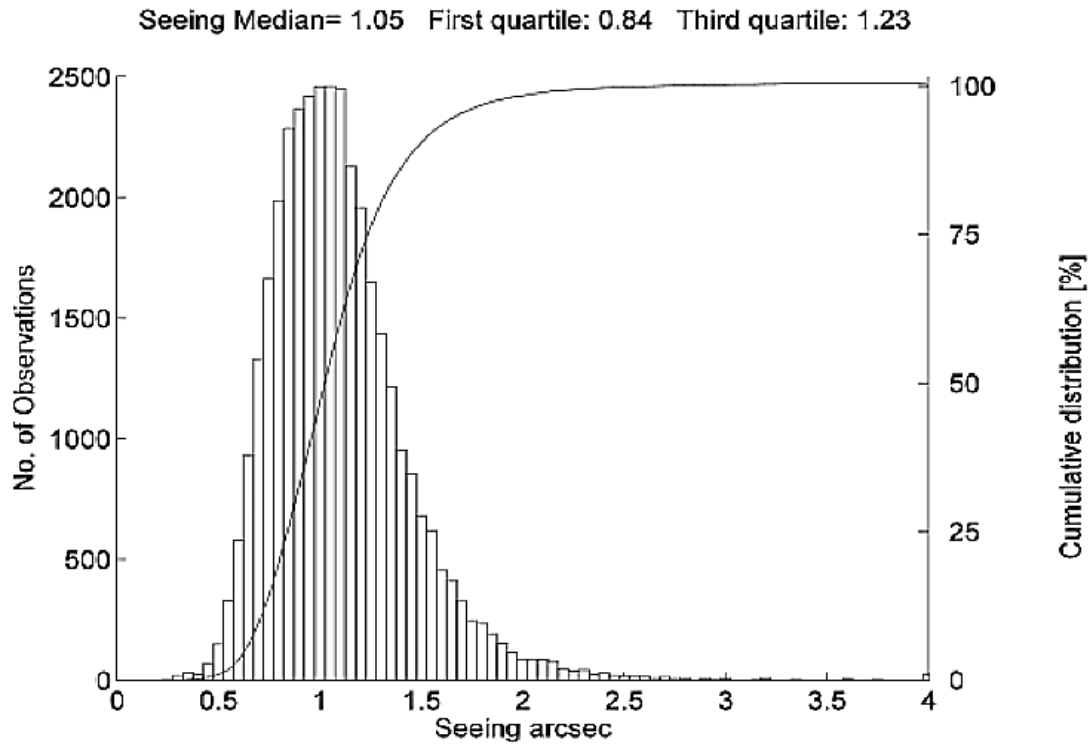

Fig.4d) Seeing histogram of Marzi site for 23 months.

## 4. Results and discussions

In addition to seeing measurements, various parameters such as sky brightness, inversion layers, precipitable water vapor, annual amount of dusty days, and relative humidity for the 4 candidate regions are investigated.

The natural sky brightness near the zenith for V, B and R bands, (Cinzano, 2000), are calculated to be 21.6, 22.4 and 20.5 mag/arcsec$^2$, respectively. However, the measured sky brightness was greater. A key factor in increasing the sky brightness is the artificial sources of urban lights, namely light pollution. To measure the real sky brightness we used the method proposed by Garstang (1991). The measurements were performed using the CCD camera. The data was collected on several occasions with no moonlight and Landolt (1992) reference stars with the magnitudes greater than 10 were used for calibration. The data was reduced using IRAF package and the results are collected in Hoseini & Nasiri (2006). The results were in good agreement with the empirical relationship proposed by Walker (1991) to estimate the light pollution as a function of the urban population and the distance from the populated area.

The frequency of occurrence of the inversion layer at different heights and the amount of precipitable water in 4 candidate regions are also investigated (Nasiri, 2004).

Considering various parameters discussed, a rough estimation of superiority order of the sites are quantitatively summarized in Table 3. The labels A=80, B=60 and C=40 as weighting indicators refer to relative importance of the parameter and the numbers 4, 3, 2 and 1 denote the superiority orders of the sites. (The larger number refers to better conditions).

| Qom | Kashan | Kerman | Birjand | Paramater/City |
|---|---|---|---|---|
| 2 | 4 | 1 | 3 | Precipitation(A) |
| 4 | 4 | 1 | 2 | Cloud Cover (A) |
| 2 | 3 | 1 | 4 | Relative humidity(B) |
| 3 | 1 | 2 | 3 | Topography(B) |
| 4 | 3 | 1 | 2 | Local specifications (B) |
| 4 | 3 | 1 | 2 | Seismicity (A) |
| 4 | 4 | 1 | 2 | Wind (A) |
| 4 | 3 | 1 | 2 | Inversion height (C) |
| 3 | 2 | 4 | 1 | Inversion (freq) (C) |
| 3 | 4 | 1 | 3 | Dust(B) |
| 3 | 3 | 2 | 1 | Light Pollution (A) |
| 2 | 2 | 2 | 2 | Sky brightness(B) |
| 2500 | 2420 | 1080 | 1820 | Total |

Table 3) Qom (Dinava) and Kashan (Kolahbarfi) suggesting promising sites.

Regarding the parameters listed in Table 3, Qom (Marzi) and Kashan (Kolahbarfi) sites, with slight differences, have more or less similar conditions. Remembering their seeing values (section 3), both sites were indicated as promising sites without a definite preference. The Birjand and Kerman sites lie in 3$^{rd}$ and 4$^{th}$ position, again in concertina with their poorer seeing conditions.

Thus, in 2007 we reported to the INO steering board that both of the Kolahbarfi and Dinava summits could be equally promising sites for construction of the INO. These sites, although were situated at high mountains. However, there was another mountain in the region named Gargash with summit height about 3600 m above sea level. It was about 450 m and 500 m higher than Kolahbarfi and Dinava, respectively. The problem for long term studies with this site was its sharp steepness such that it was not accessible, except by mule. To decide on the final site we tried to explore the seeing condition of the Gargash peak and compare the results with that of the either Dinava or Kolahbarfi site. Due to hard climbing conditions, the DIMM instruments which were used to measure the seeing parameter were remotely controlled after setting up on the summit. Fig. 5a. The same set of DIMM was installed in Dinava to collect seeing data simultaneously (Khoroshahi, 2011). As argued before, except for the seeing parameter (which may be different due to differing altitudes), the other parameters listed in Table. 3, are expected to be almost the same for these 3 sites as they are not located so far from each other in same region. (Fig. 5b).

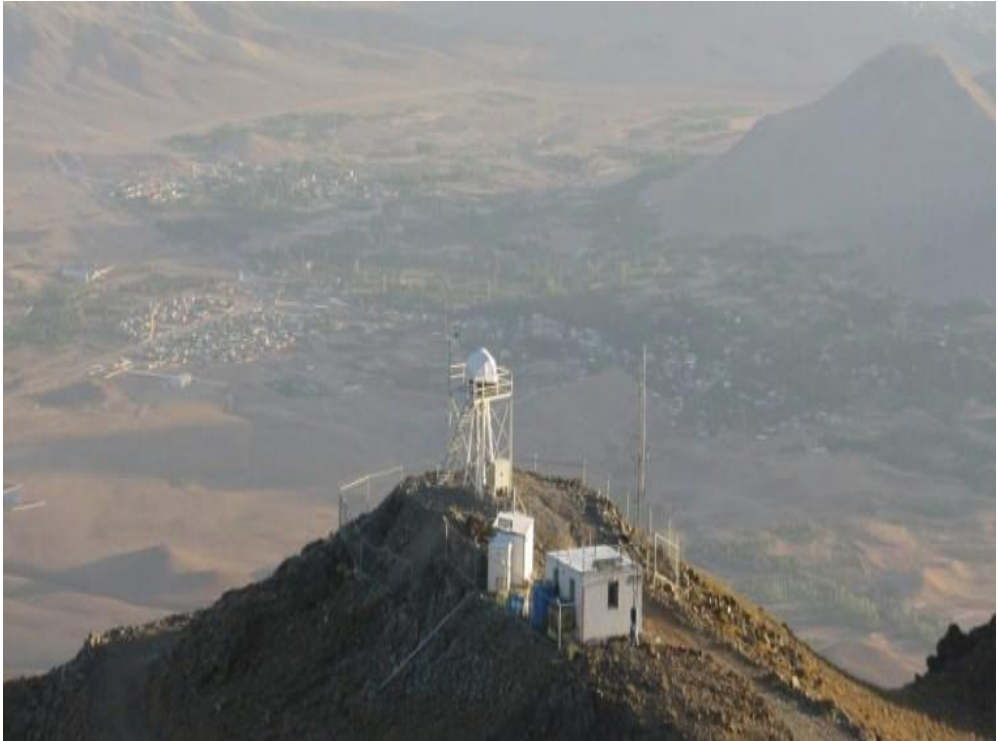

(a)

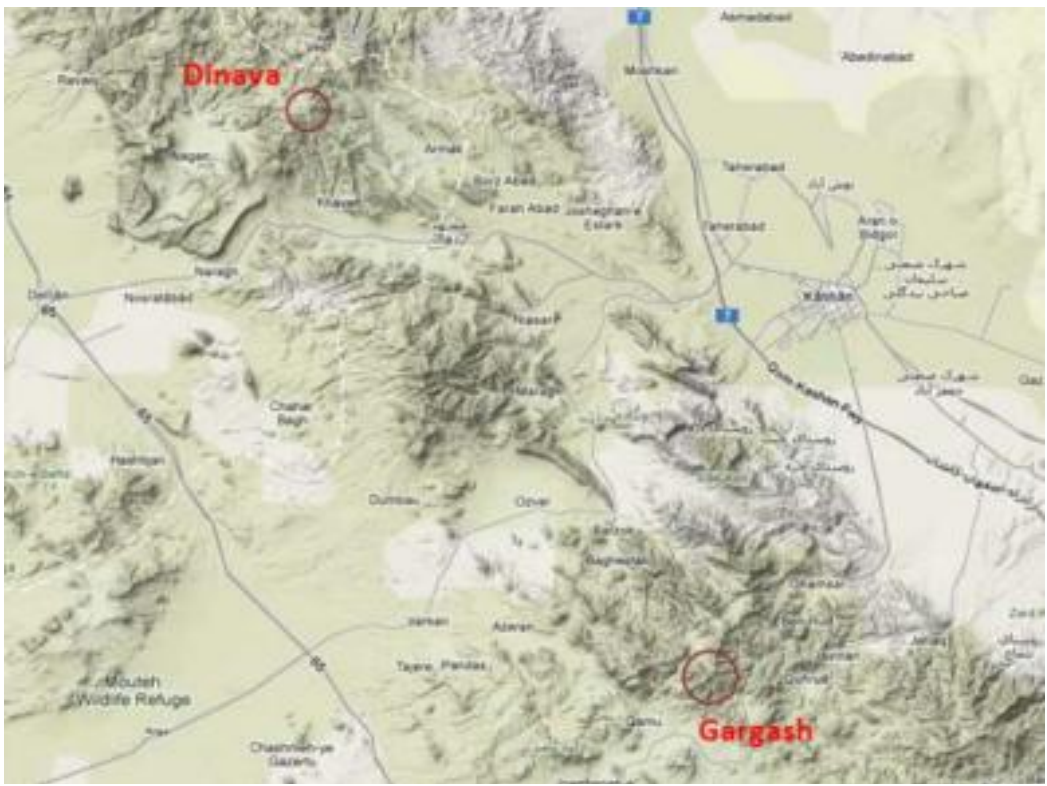

(b)

Fig. 5) A view of the (a) Gargash peak and (b) geographic location of Dinava and Gargash

A typical result obtained by further seeing measurements on these sites is shown in Fig. 6.

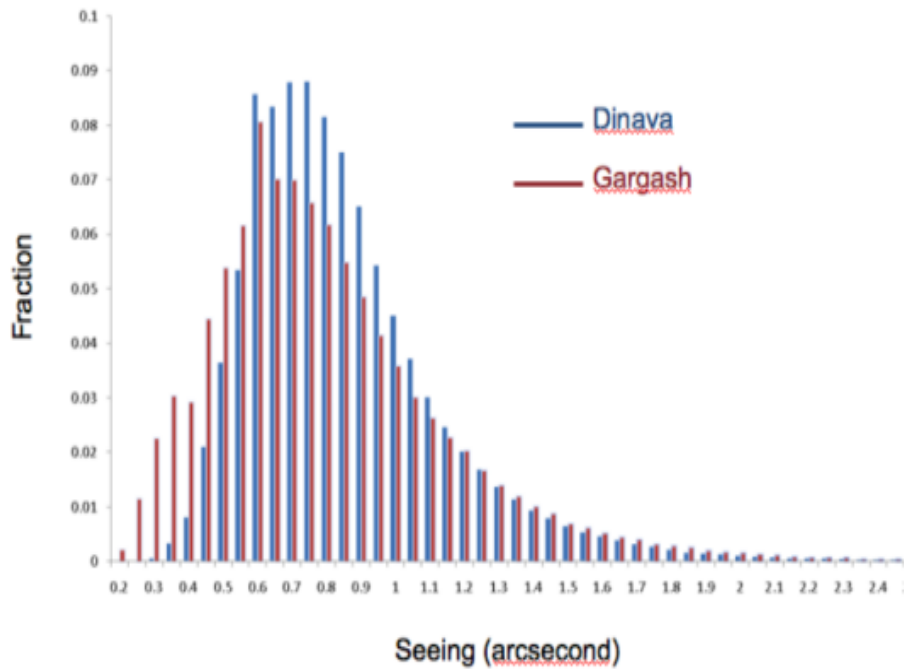

Fig. 6) Seeing comparison between Dinava and Gargesh. Data obtained for 3 months, summer 2010

As is seen from Fig. 6, as a typical seeing histogram, the median value for Dinava is $0.73''\pm0.09$ and for Gargash is $0.68''\pm0.04$, respectively. The first quartile for Divana was found to be $0.60''\pm0.09$ and for Gargash was $0.54''\pm0.04$. Therefore, the difference between seeing parameters as a major indicator for Dinava and Gargash was so considerable that the Gargash peak was selected as the final site for the construction of INO in 2010. For comparison of the other parameters leading to the selection of Mount Gargash see Khosroshahi et al (2016) and Khosroshahi (2018).

**5. Conclusions**

The Iranian National Observatory site selection was initiated in 2000 and concluded in 2010. Over a whole decade, we progressed from learning about astronomical site selection, to choosing a quality site at an altitude of 3600m which is now known as the site of the INO340 3.4m optical telescope. Meteorological data was gathered from archives and analysed for this propose. Sites visits and short term atmospheric turbulence measurements set the stage for a two-years long simultaneous and continuous seeing measurements using DIMM systems on 4 most promising sites in Iran. The final decision lay between two nearby sites of Dinava (3000m) and Gargash (3600m) after which Mount Gargash was found to be the superior because of the better seeing, higher altitude, suitable wind direction and speed and night darkness. The construction of the site is currently ongoing.

*Acknowledgements*: We would like to thank Prof. Y. Sobouti, Head of the Institute for Advanced Studies in Basic Science (IASBS) and the member of the INO coordinating committee and Prof. R. Mansouri the former manager of the INO project for their kind financial and administrative supports. We would like to express our deep gratitude to Prof. F. R. Querci form Midi Pyrenees Observatory, Prof F. Vakili from Nice University of France for their valuable scientific consultations. We are very grateful to Prof. J. Ghanbari, Proff. A. Kyasatpoor, Dr. S. Rahvar and Dr. Y Abedini for their useful joint discussions. We are also grateful to Dr. J. Fatemi, Mr. J. Noohi Mr. A. Esfahani, Dr. M. Mahmoodi, Mr. E. Hashemzeie, Dr. M. Behdani, the heads of the site selection groups in Kerman, Kashan, Qom and Bijand and Mr. A. Behnam, Mr. M. Rezvani, Mr. A. Armin, Mr. B. Sohrabi, Mr. A. Ghanbarzadeh, Mr. M. Nasiri, Mr A. Amirzadeh and Mr. A. Karimi the members of the site selection teams in the above 4 sites. S. Nasiri really appreciates the hard work of Mr. A. Abedini, Miss. S.S. Hosseini, Mr. A. Sojasi, Mrs. A. Yaghooti and Miss L. Rafezi as MSc. Students in Dept. of Physics of Zanjan Univ. doing their research project on observatory site selection under Nasiri's supervision and exploring the physical behavior of some astro-climate parameters. We appreciate the kind efforts of Mr. A. Bigdeli (IASBS Driver) and Mrs. M. Ghofrani (IASBS Secretary). We would like to express our warm thanks to all regional directors of the Universities, Institutes and the Organizations who have effectively helped us in Kerman, Kashan, Qom and Birjand regions. There were so many people that we could not put all their names here, but it does not diminish their useful assistance and guidance that we wish to warmly appreciate.